\documentclass[a4paper,12pt]{article}
\usepackage{epsf}

\title{Distortions of the Harrison-Zel'dovich 
spectrum from the QCD transition}
\author{Dominik J. SCHWARZ\thanks{Talk given at the Third International 
Conference on Physics and Astrophysics of the Quark Gluon Plasma,
March 17 -- 21, 1997, Jaipur, India.}, C. SCHMID, and P. WIDERIN \\
        Theoretische Physik, ETH-H\"onggerberg, CH-8093 Z\"urich}
\date{}

\begin{document}
\maketitle
\vspace{-7cm}
\hfill ETH-TH/97-13\\
\vspace{+7cm}

\begin{abstract}
We investigate the effect of the cosmological QCD transition
on the evolution of primordial density perturbations. If the 
phase transition is first order, the sound speed vanishes during 
the transition and density perturbations fall freely. The 
Harrison-Zel'dovich spectrum of density fluctuations
develops large peaks on scales below the Hubble radius at the transition. 
These peaks above the primordial spectrum 
grow with wavenumber and produce cold dark matter clumps of masses 
less than $10^{-10} M_\odot$. At the horizon scale the amplification
of overdensities is of order unity, and thus no $1 M_\odot$ 
black hole production is possible for a COBE normalized scale-invariant 
spectrum.
\end{abstract}

\noindent {\bf Introduction.}
The QCD transition from a hot plasma of deconfined quarks and 
gluons to a hot gas of hadrons happens in the early Universe
at a temperature $T_\star$. Lattice QCD results for the physical values
of the quark masses indicate that the QCD phase transition is 
of first order \cite{Iwasaki} and takes place at $T_\star \sim 150$ 
MeV \cite{MILC}. The QCD transition may have important 
cosmological consequences: it could lead to inhomogeneous nucleosynthesis 
\cite{Schramm}, produce dark matter in the form of strangelets \cite{Raha},
or give rise to gravitational waves. These effects have in common that 
their scale is set by the mean bubble nucleation distance $R_{\rm nucl}$ 
and not by the Hubble scale $R_{\rm H}$. It has turned out that $R_{\rm nucl} 
\ll R_{\rm H}$, mainly because the surface tension is very small \cite{ls}. 
In contrast, we study the scales $\lambda \gg R_{\rm nucl}$.

Cosmological density perturbations are affected by the QCD transition
at scales $\lambda \stackrel{<}{\scriptstyle\sim} R_{\rm H}$ 
\cite{SSW,ASC}, where $R_{\rm H} \sim m_{\rm P}/T^2_\star \sim 10^4$ m. 
For a first order QCD phase transition the
deconfined and confined phases can coexist at the coexistence
temperature $T_\star$ at fixed pressure $p_\star = p(T_\star)$. 
During this coexistence period pressure gradients, and thus the 
sound speed, vanish for wavelengths $\lambda \gg R_{\rm nucl}$. 
The vanishing sound speed gives rise to large peaks and dips above
the primordial spectrum of density perturbations. These peaks grow at 
most linearly with wavenumber. We show that the formation of black holes at 
the horizon scale is impossible for standard inflationary
scenarios, in contrast to recent claims \cite{Schramm,J}. 
We predict clumps with $M < 10^{-10}M_\odot$ in cold dark matter (CDM), 
if the CDM is kinetically decoupled at the QCD transition.
\medskip

\noindent {\bf The cosmological QCD transition.}
The QCD phase transition starts with a short period ($10^{-4} t_{\rm H}$)
of tiny supercooling, $1-T_{\rm sc}/T_\star\sim 10^{-3}$. 
When $T$ reaches $T_{\rm sc}$, bubbles nucleate at mean distances 
$R_{\rm nucl} \stackrel{<}{\scriptstyle\sim} 2$ cm \cite{CM}. The bubbles 
grow most probably by weak deflagration \cite{I}. The released energy is
transported into the deconfined phase by shock waves, which reheat
the deconfined phase to $T_\star$ within $10^{-6} t_{\rm H}$. 
The entropy production during 
this very short period is ${\rm d} S/S \sim 10^{-3}$. Further bubble 
nucleation in the reheated quark-gluon phase is prohibited after this 
first $10^{-4} t_{\rm H}$. Thereafter, the bubbles grow adiabatically
due to the expansion of the Universe. The transition completes after
$10^{-1} t_{\rm H}$.

During this equilibrium evolution the rate at which the quark-gluon 
phase can be converted into the hadron phase and vice versa,
$\Gamma_{\rm conv}$, must be compared to the Hubble rate $H$ and to the
wave number $k^{\rm phys}$ of the considered density
perturbation. $\Gamma_{\rm conv}$ cannot 
enormously differ from the typical QCD scale fm$^{-1}$, while
$H\approx (10^4$ m$)^{-1}$, therefore $H/\Gamma_{\rm conv}$ and
$k^{\rm phys}/\Gamma_{\rm conv}$ are of order $10^{-19}$. 
This means that the QCD relaxation times are incredibly 
short and that the 2-phase system is very close to equilibrium.
Since all interaction rates $\Gamma_{\rm strong,elweak}
\gg H$, all particles are in chemical and thermal equilibrium. At scales
$\lambda > R_{\rm nucl}$ we may describe 
photons, leptons, and the QCD 
matter by a single radiation fluid with equation of state $p = p(\rho)$.

During this reversible coexistence period (isentropic) $T$ and $p$ are fixed,
while $\rho$ can vary, therefore 
\begin{equation}
\label{cs2}
c_s^2 \equiv \left(\partial p\over \partial
\rho\right)_{\rm isentropic} = 0 \ .
\end{equation}
In \cite{NY} it was claimed that the isentropic condition is not consistent.
This is wrong, because they apply non-relativistic ($v \ll 1, p\ll\rho$)
hydrodynamical equations to relativistic wave fronts
in the radiation fluid. Moreover, they claim that sound propagation keeps 
the energy fraction in each phase fixed. This is incompatible with the 
equilibrium situation described above.

The thermodynamics of the cosmological QCD transition can be 
studied in lattice QCD, because the baryochemical potential is 
negligible in the early Universe, i.e.\ $\mu_{\rm b}/T_\star \sim 10^{-8}$. 
We fit the equation of state from lattice QCD \cite{Karsch}.
In addition we use the bag model to
illustrate the origin of the large peaks in the density spectrum.
\medskip

\noindent {\bf The evolution of density perturbations.}
During an inflationary epoch in the early Universe density perturbations 
$\delta\!\rho$ have been generated with a scale-invariant Harrison-Zel'dovich 
(HZ) spectrum. During the radiation dominated era subhorizon ($k^{\rm phys}
\gg H$) density perturbations oscillate as acoustic waves, supported 
by the pressure $\delta p = c_s^2 \delta\!\rho$. It is useful to study 
the dimensionless density contrast $\delta \equiv \delta\!\rho/\rho$
as a function of conformal time and comoving wave number. The subhorizon 
equation of motion reads
\begin{equation}
\label{delta}
\delta'' + c_s^2 k^2 \delta = 0 \ ,
\end{equation}
if we assume that the phase transition is short compared to the Hubble time.
From the lattice QCD equation of state the period of vanishing
$c_s^2$ lasts for $0.1 t_{\rm H}$, in the bag model it lasts 
for $0.3 t_{\rm H}$.

\begin{figure}
\begin{center}
\epsfig{figure=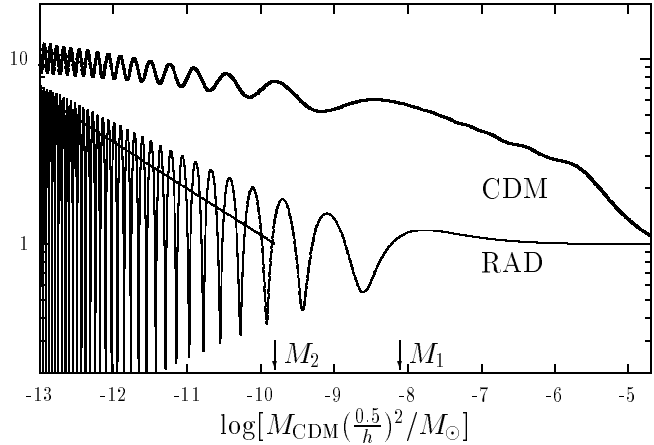,width=0.65\linewidth}
\end{center}
\vspace{-24pt}
\caption{
The modifications of the CDM density contrast $|\delta_{\rm CDM}|
(T_\star /10)$ and of the radiation fluid amplitude $A_{\rm RAD}$ 
due to the QCD transition (lattice QCD fit). Both quantities are
normalized to the pure Harrison-Zel'dovich radiation amplitude. On the
horizontal axis the wavenumber $k$ is represented by the CDM mass contained
in a sphere of radius $\pi/k$.}
\label{fig2}
\end{figure}

To illustrate the mechanism of generating large peaks above the HZ spectrum 
we discuss the subhorizon evolution of the density contrast $\delta$ in the
bag model \cite{ASC}.
Before and after the QCD transition $c_s^2 = 1/3$ and the density contrast 
oscillates with constant amplitudes $A^{\rm in}$ resp.~$A^{\rm out}$. 
At the coexistence temperature $T_\star$ the pressure gradient, i.e.\
the restoring force in in Eq.~(\ref{delta}), vanishes and
the sound speed (\ref{cs2}) drops to zero. 
Thus $\delta$ will grow or decrease linearly, depending on the fluid velocity 
at the moment when the phase transition starts. This produces large peaks 
in the spectrum which grow linearly with the wave number, i.e.\ 
$A^{\rm out}/A^{\rm in}|_{\rm peaks} \approx k/k_1$. The radiation in a 
Hubble volume at $T_\star$ has the mass $M_{\rm RAD}(R_{\rm H}) \sim 
1 M_\odot$. The CDM mass is related to the radiation mass by
$M_{\rm CDM} = a_\star/a_{\rm eq} M_{\rm RAD} \sim 10^{-8} M_{\rm RAD}$.
CDM falls into the gravitational potential wells of the radiation 
fluid during the coexistence regime. 

Fig.~\ref{fig2} shows the final spectrum for the density contrast
of the radiation fluid and of CDM. It has been obtained from numerical 
integration of the fully general relativistic equations of motion and the 
lattice QCD equation of state \cite{SSW}. The mass $M_1$ corresponds to
the wave number $k_1$ of the bag model analysis. $M_1$ 
coincides with the mass inside the Hubble horizon. The mass $M_2$ stems from
a WKB analysis with the lattice QCD equation of state, which shows that
$A^{\rm out}/A^{\rm in}|_{\rm peaks} \approx (k/k_2)^{3/4}$.  
We conclude that at the horizon scale the modification of the HZ
spectrum is mild, whereas for scales much smaller than the horizon 
big amplifications are predicted.  Without tilt in the COBE normalized
spectrum the density contrast grows 
nonlinear for $k^{\rm phys}/H \stackrel{>}{\scriptstyle \sim}
10^4$ resp.~$10^6$ for the bag model resp.~lattice QCD equation of state. 
\medskip

\noindent {\bf Implications of the large peaks.}
It was suggested that the QCD transition could lead to the formation of
$1 M_\odot$ black holes that could account for all missing (dark) matter
\cite{Schramm,J}, i.e.\ all the radiation mass
within one Hubble horizon should collapse to a single black hole. 
In common inflationary scenarios the density contrast is of order $10^{-4}$
on subhorizon scales before the QCD transition. Thus, our linear
analysis applies. From Fig.~\ref{fig2} it is clear that there is only 
a small amplification of order unity at the horizon scale (around
$M_1$). Therefore, with COBE normalized scale-invariant spectrum (allowing
tilts $|n-1| <0.3$) it is impossible to form black holes at the QCD 
transition that are abundant enough to close the Universe. 

At $T\sim 1$ MeV the neutrinos decouple from the Hubble scale,
from smaller scales they decouple at higher temperatures. 
During their decoupling they damp all
fluctuations in the density contrast due to collisional damping. Thus, at the
time of big bang nucleosynthesis all peaks of Fig.~\ref{fig2} are erased
and the energy density on scales $\lambda \stackrel{<}{\scriptstyle \sim}
R_{\rm H}(1$ MeV$)$ is homogeneous.

Weakly interacting dark matter, like the neutralino, does not belong to 
our CDM, because it is not kinetically decoupled at the QCD transition. 
For our purpose the neutralino would belong to the radiation fluid and
subhorizon perturbations in its density are damped during
its kinetic decoupling. 

However, collisional damping is irrelevant for other CDM like primordial 
black holes. Peaks in this CDM 
survive and grow logarithmically during the radiation era. Shortly after 
equality they grow nonlinear and collapse by gravitational virialization 
to clumps of $M < 10^{-10} M_\odot$. Their size can be estimated to be a 
few AU. Whether such CDM clumps may be observable remains to be 
investigated. 

D.~J.~S.~would like to thank the organizers of this conference for 
making stimulating talks and discussions possible.
D.~J.~S. and P.~W. are supported by the Swiss National Science Foundation.

\end{document}